\documentclass[twocolumn,reprint,nofootinbib,longbibliography,prd]{revtex4-1}
\usepackage{amsfonts, amsmath, amssymb, bm, enumerate, graphicx, graphics, color,mathrsfs,hyperref,nicefrac}

\newcommand{\Lie}[0]{{\cal L}\, }

\newcommand{\rIH}{r_{\mbox{\tiny{IH}}}}

\newcommand{\tl}{\theta_{(\ell)}}
\newcommand{\tn}{\theta_{(n)}}

\newcommand{\tL}{\theta_{(L)}}
\newcommand{\tN}{\theta_{(N)}}

\newcommand{\be}{\begin{equation}}
\newcommand{\ee}{\end{equation}}
\newcommand{\bea}{\begin{eqnarray}}
\newcommand{\eea}{\end{eqnarray}}

\newcommand{\tq}{\tilde{q}}

\newcommand{\tom}{\tilde{\omega}}

\begin{document}

\title{	
Unstable marginally outer trapped surfaces in static spherically symmetric spacetimes}

\date{\today}
\author {Ivan Booth}
\email{ibooth@mun.ca}
\affiliation{
Department of Mathematics and Statistics\\ Memorial University of Newfoundland \\  
St.~\!\!John's, Newfoundland and Labrador, A1C 5S7, Canada \\
}
\author{Anna O'Grady}
\email{ogrady@astro.utoronto.ca}
\affiliation{
Department of Physics and Physical Oceanography\\ Memorial University of Newfoundland \\  
St.~\!\!John's, Newfoundland and Labrador, A1C 5S7, Canada \\
}
\affiliation{Dunlap Institute for Astronomy and Astrophysics\\ Department of Astronomy and Astrophysics \\  University of Toronto\\
50 George Street, Toronto, Ontario, M5S 3H4, Canada}
\author{Hari K. Kunduri}
\email{hkkunduri@mun.ca}
\affiliation{
Department of Mathematics and Statistics\\ Memorial University of Newfoundland \\  
St.~\!\!John's, Newfoundland and Labrador, A1C 5S7, Canada \\
}

\begin{abstract}
We examine potential deformations of inner black hole and cosmological horizons in Reissner-Nordstr\"om de-Sitter spacetimes. While the rigidity of 
the outer black hole horizon is guaranteed by theorem, that theorem applies to neither the inner black hole nor past cosmological horizon. Further for  
 pure deSitter spacetime it is clear that the cosmological horizon can be deformed (by translation). For specific parameter
choices, it is shown that both inner black hole and cosmological horizons can be infinitesimally deformed. 
However these do not extend to finite deformations. 
The corresponding results for general spherically symmetric spacetimes are considered.
\end{abstract}

\maketitle

\section{Introduction}

In stationary spacetimes, the event horizon of a black hole is a Killing horizon and foliated by surfaces with vanishing outward null expansion: 
marginally outer trapped surface (MOTS). 
More generally given a Cauchy surface in any spacetime, the boundary of the trapped region is an apparent horizon which is also a MOTS. 
Motivated by these facts MOTS are key to many definitions of black hole boundaries including trapping horizons\cite{Hayward:1993wb}, 
marginally trapped tubes\cite{Ashtekar:2005ez}, isolated and dynamical horizons\cite{Ashtekar:2004cn}, the proposed core of the trapped 
region\cite{Bengtsson11} and the very recent future holographic screens\cite{Bousso:2015mqa}. 

Apart from being foliated by MOTS the event horizons of the standard stationary black hole solutions (for example Kerr-Newman-deSitter) 
have another property: they separate the trapped region from the untrapped region and in particular there are fully trapped surfaces
uniformly close to and  ``just inside'' the MOTS. MOTS with slight variations of this property go by many names including \emph{stable}\cite{Newman87},  \emph{outer trapping}\cite{Hayward:1993wb} or \emph{strictly stably outermost}\cite{Andersson:2005gq}. 
MOTS with one of these properties and which foliate a stationary event horizon have been shown to be geometrically rigid against deformations\cite{Andersson:2005gq,Booth07}. 

\begin{figure}
\includegraphics[scale=1]{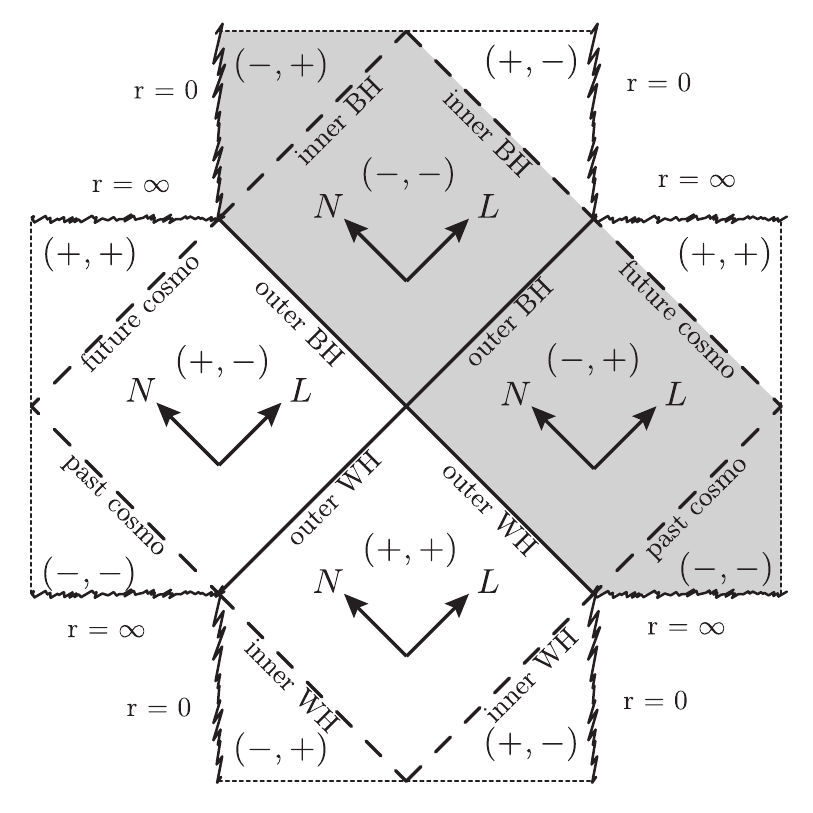}
\caption{One tile of the Penrose-Carter diagram for Reissner-Nordstr\"om-deSitter spacetime. It repeats (subject to possible identifications) across 
the dotted lines. The various Killing horizons are labelled: outer and inner black hole, outer and inner white hole, and future and past cosmological 
horizons. The null directions are consistently labelled as $L$ and $N$ with the signs of $\tL$ and $\tN$ in each region being respectively listed as $(\pm,\pm)$. $\tL$ and $\tN$ vanish on horizons to which they are tangent. Stable horizons are solid lines while unstable  are dashed. 
Most of our discussion will focus on the shaded region containing outer and inner black hole horizons along with the past-cosmological horizon.  
}
\label{RNdSHorizons}
\end{figure}

However even in stationary spacetimes stable horizons are the (admittedly very important) exception rather than the rule. 
Consider, for example, Reissner-Nordstr\"om-deSitter as depicted in Figure \ref{RNdSHorizons}. Focussing on the shaded region, 
the inner black hole and past cosmological horizons have trapped surfaces outside rather than inside and so are not stable\footnote{
The terminology of outside versus inside becomes ambiguous in some of these cases but we will return to clarify this in Section \ref{horizons}.
}.


Further, there is at least one case where an unstable horizon \emph{can} be smoothly deformed. Pure deSitter spacetime is homogeneous and isotropic
with a constant positive Ricci curvature $\mathcal{R} = 4\Lambda$ determined by the cosmological constant. However around any point $p$ in 
the space it is possible to construct the standard static coordinate patch:
\be
ds^2 = - \left(1-\frac{\Lambda}{3}r^2 \right) dt^2 + \frac{dr^2}{1-\frac{\Lambda}{3}r^2} + r^2 d \Omega^2
\ee
from which it is straightforward to show that there is a MOTS at 
\be 
r_{\mbox{\tiny{CH}}} =  l \equiv \sqrt{\frac{3}{\Lambda}} \; . 
\ee 
Like the cosmological horizons in RNdS, it isn't  stable and it is intuitively clear that in this case we can deform the MOTS. To see this construct
an analogous MOTS around a point $p'$ infinitesimally close to $p$. This represents a deformation of the $p$-MOTS which in this case
is essentially a translation. 

So at least in this case the lack of stability corresponds to a freedom to deform the MOTS. In the rest of this paper we will explore unstable
MOTS in more detail. Section \ref{background} reviews basic nomenclature and geometry along with the standard stability result. 
Section \ref{InfDef} shows that only particular finely tuned unstable MOTS in RNdS can be infinitesimally deformed. Section \ref{FinDef}
examines whether these infinitesimal deformations can be made finite (and so real!). Section \ref{Discuss} summarizes and discusses our results.
Appendix \ref{triple} reviews some useful identities for Legendre polynomials that are applied in the main text.

\section{Background and general theory}
\label{background}

We begin with a very brief review of the mathematics and geometry of marginally outer trapped surfaces and their deformations. 
As general references for the next two subsections, see \cite{Booth07} for more details on the geometry or \cite{Booth:2005qc} for a review of the various types MOTS and their complications. 

\subsection{Spacetime and two-surface geometry}
\label{geometry}

Let $(M, g_{ab}, \nabla_a)$ be a $(3+1)$-dimensional time-oriented spacetime and $(S, \tq_{ab}, d_a)$ be a spacelike 
closed and orientable two-surface  embedded in $M$.

The normal space at each $p\in S$ is two-dimensional and timelike and so can be spanned by a pair of null vectors. In particular since $M$ is
time-oriented we may define a pair of future-oriented vector fields $\ell$ and $n$ over $S$ which do this job in each normal space. 
Since they are null vector fields they each have one-degree of rescaling freedom. However one of these is removed by requiring 
that they be cross-scaled so that $\ell \cdot n = -1$.


Geometric consistency requires that the combined induced metric/projection operator on $S$ satisfies 
\be
\tilde{q}_{ab} = g_{ab} + \ell_a n_b + n_a \ell_b  \, .  \label{tq}
\ee
This fixes the intrinsic geometry of $S$ while the extrinsic geometry comes from 
tangential derivatives of the null-normals. These are the connection on the normal bundle
\be
\tilde{\omega}_a = -\tq_a^c n_b  \nabla_c \ell^b \, . 
\ee
and the extrinsic curvatures
\be
k^{(\ell)}_{ab} = \tq_a^c \tq_b^d \nabla_c \ell_d  \; \; \mbox{and} \; \; k^{(n)}_{ab} =  \tq_a^c \tq_b^d  \nabla_c n_d  \, ,
\ee
which may be conveniently decomposed into their trace and trace-free parts
\be
k^{(\ell)}_{ab} = \frac{1}{2} \theta_{(\ell)} \tq_{ab} + \sigma^{(\ell)}_{ab}  \; \; \mbox{and} \; \;  k^{(n)}_{ab} = \frac{1}{2} \theta_{(n)} \tq_{ab} + \sigma^{(n)}_{ab} \; . 
\ee
Respectively these are the expansions and the shears of those vector fields. The reason for these names is clear if we consider their alternate
definition as deformations (also known as variations\cite{Andersson:2005gq}). 

Consider a vector field $X^a$ which is normal to $S$ and defined in a neighbourhood of the surface. Hence on $S$
\be
X^a = A\ell^a - B n^a \label{X0}
\ee
for some functions $A$ and $B$.\footnote{The negative sign for $B$ is a convention chosen in \cite{Booth07} to simplify calculations when 
studying dynamical horizons. Even though they are not considered here, we retain the sign for consistency with that paper from which we draw 
almost all of our formulae.} Then that function defines a flow which can be used to evolve and deform $S$.

In coordinate terms, if $x_{S}^\alpha (\theta,\phi)$ is a (local) parameterization of $S$ then infinitesimally the deformation sends:
\be
x_{S}^\alpha (\theta,\phi)  \rightarrow x_{S}^\alpha (\theta,\phi) + \varepsilon X^\alpha(\theta,\phi) \, .
\ee
The evolution also identifies points on the original and deformed surfaces along the lines of flow. Thus one may consider the
rate of change of the geometric properties of the surface under the deformation and we denote this differential operator as $\delta_X$. It
is usually referred to as either the \emph{deformation operator} or \emph{variation} with respect to $X$. 

If the coordinate system is adapted to $S$ and $X^a$ so that $S$ is a level surface and $X=\partial/\partial \lambda$ a coordinate 
vector field then $\delta_X$, the Lie derivative $\Lie_X$ and the partial derivative $\partial/\partial \lambda$ are all the same thing. 
This equivalence is often used to simplify discussions of deformations in spherical symmetry. See, for example \cite{Hayward:1993wb}.

It is not hard to show that
\be
\delta_X \tilde{q}_{ab} = A k^{(\ell)}_{ab} - B k^{(n)}_{ab} \label{deltaq}
\ee
whence 
\be
\delta_X \tilde{\epsilon}_{ab} = ( A \theta_{(\ell)} - B \theta_{(n)} ) \tilde{\epsilon}_{ab} \, 
\ee
where $\tilde{\epsilon}$ is the area two-form on $S$ (in coordinate form $\tilde{\epsilon} = \sqrt{\tilde{q}} d\theta \wedge d\phi$). Thus 
$\theta_{(X)}$ and $\sigma^{(X)}_{CD}$ are respectively the expansion and shear of $S$ as it is evolved by $X^a$. Note too that for these 
intrinsic quantities the rates of change are independent of how $X^a$ extends off $S$. 

One can also calculate variations of the extrinsic quantities,  
however for our purposes we only need
\begin{align}
 \delta_X \theta_{(\ell)}  
&  =  \kappa_X \theta_{(\ell)}   - d^2 B + 2 \tilde{\omega}^a d_a B  \label{deltatl} \\
& - B \left(-d_a \tilde{\omega}^a + \| \tom \|^2  - \tilde{K} + G_{ab} \ell^a n^b - \tl \tn \right) \nonumber \\
& +A \left(-\|  {\sigma^{(\ell)}} \|^2 - G_{ab} \ell^a \ell^b  - (1/2) \tl^2 \right) \, .  \nonumber
\end{align}
Newly appearing quantities are $\kappa_X = - X^a n_b \nabla_a \ell^b$,  
 $\tilde{K}$  the Gaussian curvature of $S$  and 
$G_{ab}$ the Einstein tensor. Further we  have abbreviated $d^2 = d^a d_a$, 
$\| \tom \|^2 = \tom^a \tom_a$ and $\| {\sigma^{(\ell)}} \|^2 = \sigma^{(\ell) ab} \sigma^{(\ell)}_{ab}$. 

Unlike (\ref{deltaq}) this variation does depend on derivatives off $S$. This is through the gauge dependent $\kappa_X$ term which 
under rescalings $\ell \rightarrow e^f \ell$ and $n \rightarrow e^{-f} n$ of the null vectors transforms as
\be
\kappa_X \rightarrow \kappa_X - \Lie_X f \, . 
\ee
However, as will now be seen, we are only really interested in situations where $\tl$ vanishes and so do not need to worry about this dependence. 

\subsection{MOTS: definition, deformation and difficulties}
\label{horizons}

The standard classification of two-surfaces as trapped, untrapped or marginally trapped assumes that one can unambiguously assign
one of the null directions (say $\ell$) as outward pointing and the other ($n$) as inward pointing. Then a closed, spacelike two-surface $S$ 
is \emph{outer untrapped} if $\tl>0$, an \emph{outer trapped} if $\tl < 0$ and \emph{marginally outer trapped (MOTS)} if $\tl = 0$. 
A fully \emph{trapped} surface has both $\tl<0$ and $\tn<0$. 

A Killing horizon is null and so if it is tangent to the outgoing direction $\ell$, then any two-dimensional slice of that horizon is a MOTS. 
Thus for those MOTS
\be
\delta_\ell \tl = - \|  {\sigma^{(\ell)}} \|^2 - G_{ab} \ell^a \ell^b = 0 \, . \label{dltl}
\ee
Now, intuitively the outer black hole Killing horizon of a stationary spacetime should have outer trapped surfaces ``just inside''.  
In terms of deformations, the existence of such surfaces implies that for some inward-oriented spacelike normal vector field 
$R = \alpha \ell - \beta n$ ($\alpha \beta < 0$):
\be
\delta_R \tl = - d^2 \beta + 2 \tilde{\omega}^a d_a \beta  -\beta \delta_n \tl  < 0 \, , \label{drtl}
\ee
where 
\be
\delta_n \tl = -d_a \tilde{\omega}^a + \| \tom \|^2  - \tilde{K} + G_{ab} \ell^a n^b \, . \label{dntl}
\ee
The vanishing of $\delta_\ell \tl$ renders the value of $\alpha$ irrelevant. Note too that if we rescale the null vectors so that $n \rightarrow \beta n$ 
(and $\ell \rightarrow \ell/\beta$) this condition becomes $\delta_n \tl  < 0$.\footnote{Restricting attention spherical horizons, the correct scaling is
obvious but  for a concrete demonstration of a less trivial situation see the discussion of Kerr in appendix C of \cite{Booth07}.}

Now, a closed MOTS slice $S$ of a Killing horizon with $\delta_n \tl < 0$ is guaranteed to be geometrically stable in that it cannot be 
smoothly deformed out of the Killing horizon while preserving $\tl = 0$. To see this consider variations generated by a vector field 
$X$ of the form (\ref{X0}) with $B$ not everywhere vanishing (that would correspond to a variation along the Killing horizon). 
Then any such MOTS-preserving variation of $S$ necessarily satisfies
\be
\delta_X \tl = 0 \;  . \label{dXtL0}
\ee
However as considered above $\delta_\ell \tl = 0$ and again $A$ is irrelevant. Thus the variation must satisfy
\be
-d^2 B + 2 \tom^a d_a B - B \delta_n \tl = 0 \, .  \label{dxtleq}
\ee
For $\delta_n \tl < 0$ there are no solutions to this equation and so no MOTS-preserving variation\cite{Andersson:2005gq} (this can also be seen by 
a maximum principle argument \cite{Booth07}). 

MOTS satisfying versions of this condition have been considered many times over the 
years and among other names have been termed \emph{stable}\cite{Newman87},  \emph{outer trapping}\cite{Hayward:1993wb} or 
\emph{strictly stably outermost}\cite{Andersson:2005gq}. In this paper we will generally refer to them as stable. 

This set-up and labelling is all very well for outer black hole horizons in a spacetime with an unambiguous notion of ingoing and outgoing, however
in a multi-horizon spacetime like that shown in FIG.~\ref{RNdSHorizons}, outward and inward labels are not well-defined. Neither $L$ nor
$N$ is consistently outward-pointing (towards an $r=\infty$) or inward-pointing (towards an $r=0$). 

While there are 
systems of nomenclature that distinguish between the various types of horizons without reference to ``inner'' and ``outer''\cite{Senovilla:2007zy,Hayward:2009ji}, for this paper we will instead just abuse the name ``MOTS'' and use it to refer to any surface with one 
vanishing null expansion. We will always label that  direction $\ell$ (so that $\tl = 0$) and the other future null direction $n$. Note that the 
geometric stability arguments made in the paragraph surrounding (\ref{dxtleq}) continue to apply regardless of the orientation of $\ell$ and $n$. 
Thus we may always test the geometric stability of a MOTS by checking for a scaling of the null vectors such that $\delta_n \tl < 0$. 

Turning once again to FIG.~\ref{RNdSHorizons} we see that on some horizons $\ell = L$ while on others we will have $\ell = N$. However whatever
the labelling, the outer black and white hole horizons are stable by this measure while all cosmological and inner black hole horizons 
are potentially unstable with $\delta_n \tl > 0$. 

For the rest of this paper we will investigate whether this potential instability translates into finite MOTS-preserving variations.

%
%


\section{``Unstable'' MOTS in RNdS}
\label{InfDef}

In the last section we tested stability based on how the null expansions do or don't change signs across a horizon. However to understand 
whether the lack of a proof of stability actually corresponds to a real instability we need more calculations and commence with finding 
exact expressions for $\delta_n \tl$. 

\begin{figure}
\includegraphics[scale=0.5]{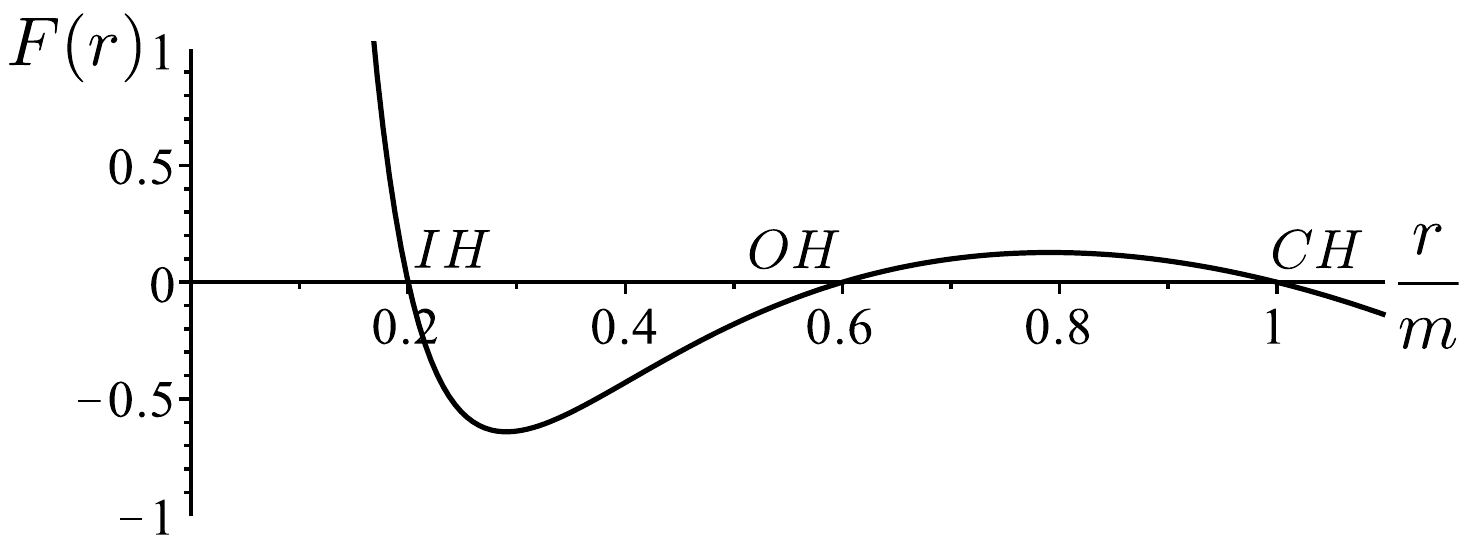}
\caption{$F(r)$ for a typical cosmological black hole solution. In this case $\Lambda \approx 0.1417/m^2$ and $q^2 \approx 0.8496 m^2$. The inner black hole horizon,
outer black hole horizon and cosmological horizon are respectively labelled as IH, OH, and CH.}
\label{Typical}
\end{figure}

\begin{figure}
\includegraphics[scale=1]{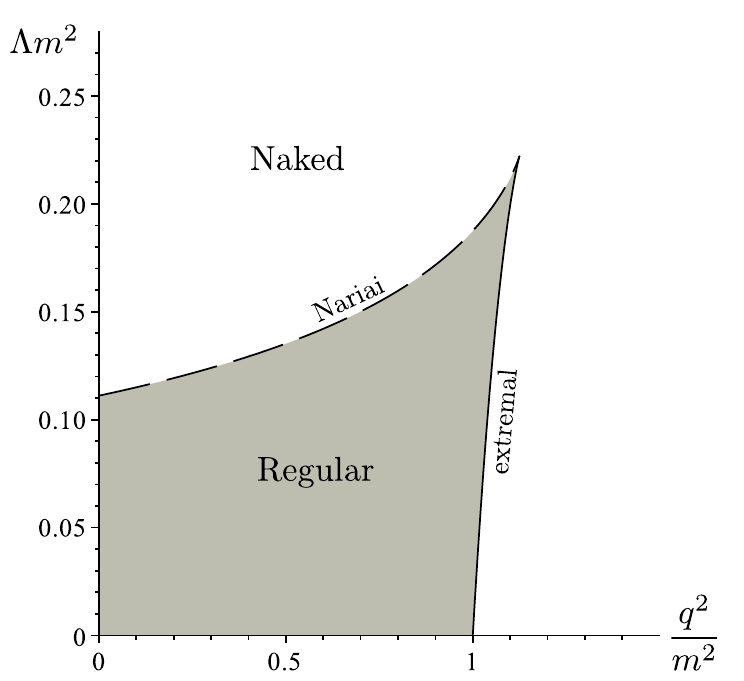}
\caption{Phase space of RNdS spacetimes with $m \neq 0$. $\Lambda$ and $q^2$ are given in units of $m$. Solutions with three horizons
(inner black hole, outer black hole, cosmological) are found in the grey shaded region while naked solutions with a single cosmological horizon form the rest
of the phase space. The only exception is pure RN along the  $\Lambda=0$ line where for $q<m$ there are inner and outer black hole horizons 
while $q>m$ is horizon-free.  Along the dashed line outer and cosmological horizons are degenerate as in the Nariai limit of SdS while the solid line
represents extremal black holes. Where the lines meet all three horizons are degenerate.    }
\label{RNdSRange}
\end{figure}

First, the RNdS metric in standard form (static for $F(r)>0$) is 
\be
ds^2 = - F(r) dt^2 + \frac{dr^2}{F(r)} dr^2 + r^2 d \Omega^2 
\ee
with
\be 
F(r) = -\frac{\Lambda}{3}r^2 + 1 - \frac{2m}{r} + \frac{q^2}{r^2} \, . \label{F}
\ee
The horizons are located at roots of $F(r)$. 
%
For black hole solutions like that depicted in FIG.~\ref{Typical} there is one negative, unphysical, root and three positive roots that 
correspond to horizons. In increasing order these are the inner black/white hole horizons at $r_{\mbox{\tiny{IH}}}$, outer black/white hole horizons
$r_{\mbox{\tiny{OH}}}$ and future and past cosmological horizons $r_{\mbox{\tiny{CH}}}$. However not all members of this family of solutions are 
cosmological black holes. FIG.~\ref{RNdSRange} shows the allowed parameter range. It was produced by examining where the discriminant of 
$F$ vanishes (these are double or triple roots and so the boundaries of the ``Regular'' region). 

For spherically symmetric $r$=constant surfaces  and a similarly symmetric scaling of the null vectors we have $\tom_a = 0$.  Thus
(\ref{dntl}) becomes 
\be
\left. \delta_n \tl\right|_{\mbox{\tiny{spherical}}} = - \frac{1}{r^2} + G_{ab} \ell^a n^b \, .
\ee
Note that this is invariant with respect to the scaling of the null vectors and in fact we can find it without ever defining them.
By (\ref{tq}),  $G_{ab} \ell^a n^b = \frac{1}{2} G_{ab} \left(\tq^{ab} - g^{ab} \right)$ so
\be
\left. \delta_n \tl\right|_{\mbox{\tiny{RNdS}}} =  -  \frac{F'}{r}  
\ee
where the prime indicates a derivative with respect to $r$ and we have applied $F(r)=0$. With $\Lambda >0$ the asymptotic behaviour is 
fixed and so the requirement that there be three positive roots means that $F(r)$ will generically take a form similar 
to FIG.~\ref{Typical}. In particular it is clear that
\be
\left. \delta_n \tl \right|_{\mbox{\tiny{OH}}} < 0 \; \; \mbox{while} \; \; \left. \delta_n \tl \right|_{\mbox{\tiny{IH,CH}}} > 0
\ee
as claimed earlier. 

However we now demonstrate that in (at least) the vast majority of cases the inner and cosmological horizons are also stable. To see this we consider
concrete solutions of the stability   equation  (\ref{dxtleq}).
On both horizons this becomes 
\be
 \nabla^2 B +   (r^2  \delta_n \tl) B = 0 \, , 
\ee
where $\nabla^2 B$ is the regular spherical Laplace operator on a unit sphere. Thus potential deformations must satisfy
\begin{align}
& \nabla^2 B = \left( r F' \right) B  \, .  \label{SL}
\end{align}
The only non-diverging solutions of the spherical Laplace eigenvalue equation are spherical harmonics. That is if there
is an integer $l$ such that 
\be
 -r F' =  l (l+1)
\ee
then (\ref{SL}) has solutions
\be
B = P_l (\cos  \theta) (A_m \cos (m\phi) + B_m \sin (m \phi) )
\ee
for integers $ 0 \leq m < l$ and constants $A_m$ and $B_m$. 

We can then test for cases where these conditions might be met. First for $m=0$ the only non-naked singularity spacetime is pure deSitter. In that case it is straightforward to 
see that $-rF'=2$ on the cosmological horizon and so the MOTS-translation freedom manifests itself as an $l=1$ instability. 

\begin{figure}
\includegraphics[scale=0.22]{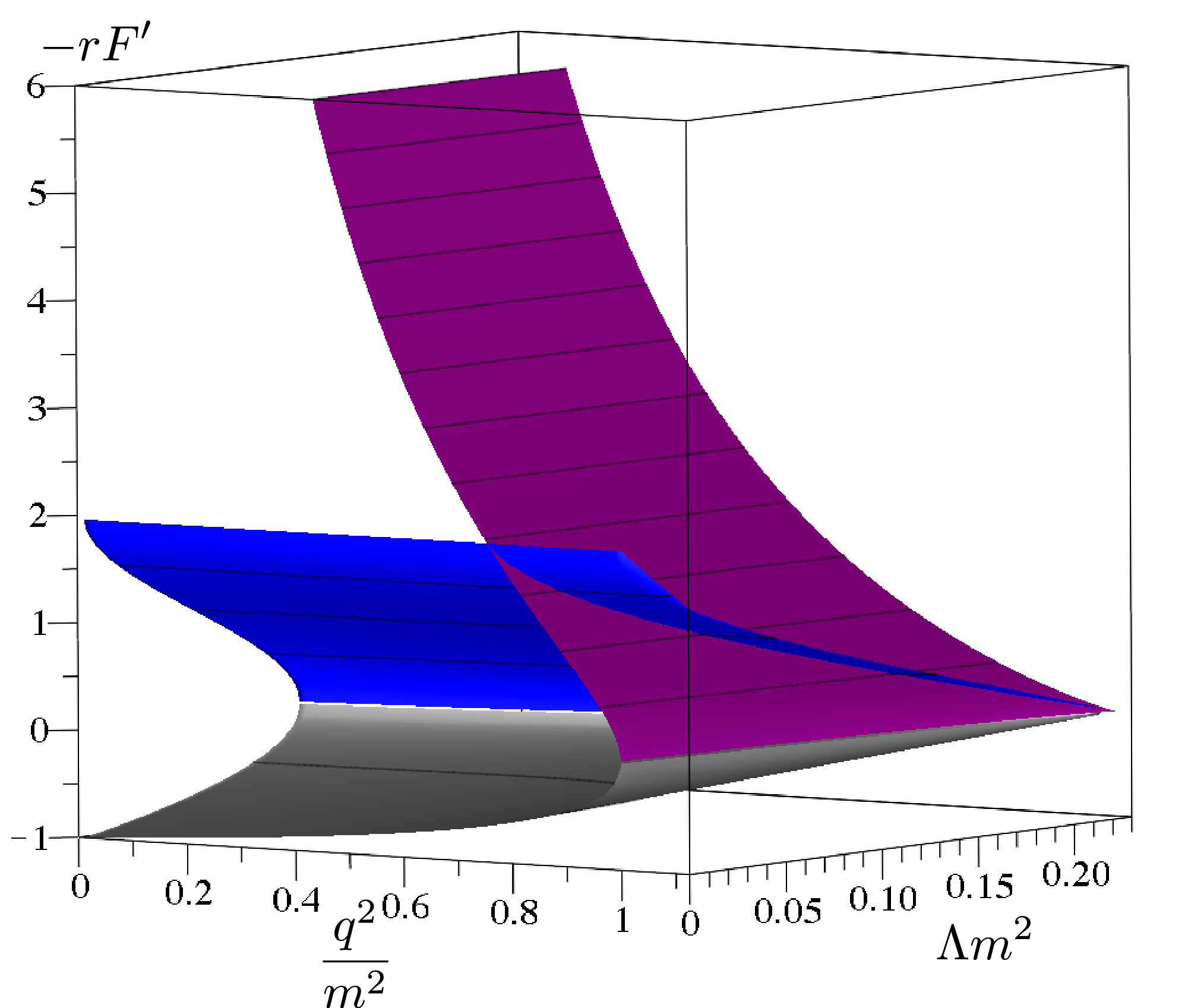}
\caption{Values of $r^2 \delta_n \theta_{(\ell)} = - r F'$ for cosmological (blue), outer (grey) and inner (purple) horizons. Potential 
instabilities exist when $- r \frac{dF}{dr} = l (l+1)$ for some positive integer $l$. Though it is cut off in the figure, the inner horizon sheet 
diverges to infinity. The domain for all horizons is as shown in FIG.~\ref{RNdSRange}.}
\label{Degree}
\end{figure}

Turning to $m \neq 0$, FIG.~\ref{Degree} shows the values $- r F'$ for all horizons in RNdS black hole
spacetimes and so  we can consider them case-by-case. 

First for the outer horizon $-1 \leq -r F' \leq 0$ and so there are no possible solutions. This is not a surprise as we have already twice 
concluded that outer black and white hole horizons are stable. 

Similarly simple is the inner horizon with $0 \leq -r F' < \infty$. In this case a correct choice of parameter values will allow any possible $l$.
In particular this is even possible for pure RN. 
 
The cosmological horizon is a little more subtle. The figure shows that the only possible case is $-rF' =2 \Leftrightarrow l=1$ 
however this limit isn't achieved: it is along the $\Lambda = 0$ line where there isn't a cosmological horizon. 
So for $m \neq 0$ there are no solutions and the cosmological horizon is stable.


Thus we have now explicitly demonstrated that while the stability condition $\delta_X \tl <0$ may be sufficient to exclude deformations it certainly isn't necessary. 
Examples are the cosmological horizon in $m \neq 0$ RNdS spacetimes and (at least) all but a finely tuned set of inner 
horizons.  In the next section we will further examine those special cases.


\section{Finite Deformations in RNdS}
\label{FinDef}

In this section we develop the formalism necessary to test the higher order stability of the special cases that were found to be first order unstable in the previous section. 
It is easiest to do this by moving away from the general formalism of Section \ref{background} and to one  specialized to the RNdS spacetimes. 
In all the cases that we check we will see that the apparent first-order instability fails at higher order.


%
%
%

\subsection{MOTS in Painlev\'{e}-Gullstrand coordinates}

We begin by introducing Painlev\`{e}-Gullstrand (PG) coordinates for RNdS spacetimes. Recall that time in these coordinates is measured along
a congruence of infalling timelike geodesics while the spatial slices of constant time are intrinsically flat\cite{Martel:2000rn}. 
For RNdS the shaded patch shown in FIG.~\ref{RNdSHorizons} is (almost) covered by coordinates $(T,r,\theta,\phi)$ with metric:
\be
ds^2 = - F(r) dT^2 + 2 \sqrt{1-F(r)} dT dr + dr^2 + r^2 d \Omega^2 \, , 
\ee
where $F(r)$ takes its usual form (\ref{F}). The ``almost''  is included in the previous sentence because for $q \neq 0$ there will always be 
a region where $1-F < 0$ and so the coordinate system is not well-defined. However, as we shall see, for example in FIG.~\ref{Fin_RN}, this will always be inside
the inner horizon and so not cause us any problems. 

We will look for MOTS on hypersurfaces $\Sigma_T$ of constant $T$ and in order to do this it will be 
sufficient to know the intrinsic and extrinsic geometry of $\Sigma_T$. The intrinsic geometry on $\Sigma_T$ is given by the Euclidean metric 
\be
d\Sigma^2 = h_{ab} dx^a dx^b =   dr^2 + r^2 d \Omega^2 \, ,
\ee
while the extrinsic curvature is 
\be
K =  \left(  \frac{F'}{2 \sqrt{1-F}} \right)  dr^2 -  \left(  r \sqrt{1-F} \right)  d \Omega^2
\ee
which was calculated from the future-oriented  unit timelike normal to  $\Sigma_T$: 
\be
\hat{u} = \left( \frac{\partial}{\partial T}\right) -  \sqrt{1-F} \left(\frac{\partial}{\partial r} \right) \, . 
\ee

Then consider a rotationally symmetric surface $S$ in a $\Sigma_T$ and parameterize it by coordinates
$(\lambda, \phi)$ as
\be
(T,R,\theta,\phi) = (T_o,R(\lambda),\Theta(\lambda) , \phi) \,  ,
\ee
for some functions $R(\lambda)$ and $\Theta(\lambda)$. For now we will find it convenient to take $\lambda$ to be the 
 arclength parameter as measured from the north pole of $S$ along the constant $\phi$ lines of longitude.
Then the tangent vector
\be
\frac{d}{d \lambda} = \dot{R} \left( \frac{\partial}{\partial r}  \right) + \dot{\Theta} \left( \frac{\partial}{\partial \theta} \right) 
\ee
is unit length
\be
\dot{R}^2 + R^2 \dot{\Theta}^2  = 1 \, , \label{arc}
\ee 
where we have marked derivatives with respect to $\lambda$ with dots.  

Next the induced two-metric on $S$ is
\be
dS^2 = d \lambda^2 + (R^2 \sin^2 \! \Theta) d \phi^2 \, , 
\ee
with inverse:
\be
\tilde{q} =    \left( \frac{\partial}{\partial \lambda} \right)   \otimes \left( \frac{\partial}{\partial \lambda} \right)  
+ \frac{1}{R^2 \sin^2 \! \Theta} \left( \frac{\partial}{\partial \phi} \right) \otimes  \left( \frac{\partial}{\partial \phi} \right) \nonumber \, .  
\ee

The positive-$r$ pointing spacelike normal to $S$ in $\Sigma_T$ is 
\be
\hat{r} = R \left( \dot{\Theta} \left( \frac{\partial}{\partial r}  \right) 
 - \frac{\dot{R}}{r^2} \left( \frac{\partial}{\partial \theta}  \right)  \label{tr}
 \right) \, ,  
\ee
whence the trace of the extrinsic curvature of $S$ in $\Sigma_T$ is
\be
\theta_{(\hat{r})} \equiv \tilde{q}^{ab} \nabla_a \hat{r}_b =  - \frac{\ddot{R}}{R \dot{\Theta}} + 2 \dot{\Theta} - \frac{\dot{R}}{R} \cot (\Theta) \, , 
\ee
where we have used the arclength condition to somewhat simplify the expression. 
Note that no $F$ appears in this expression: $\Sigma_T$ is Euclidean so any geometric calculation intrinsic to $\Sigma_T$ is independent of 
$F$. 

Next the trace of the extrinsic curvature of $S$ with respect to $\hat{u}$ (and so out of $\Sigma_T$) is:
\be
\theta_{(\hat{u}) }   \equiv \tilde{q}^{ab} \nabla_a \hat{u}_b =  K_{ab} h^{ab} -  K_{ab} \hat{r}^a \hat{r}^b \, . 
\ee
That is
\be
\theta_{(\hat{u}) } = \frac{\left(R F' + 2 (1-F)\right) \dot{R}^2 - 4 (1-F)  }{2 R \sqrt{1-F}}
\ee
Then an outward oriented null vector is $\ell = \hat{u} + \hat{r}$ and if
\be
\tl = \theta_{\hat{u}} + \theta_{\hat{r}} = 0 
\ee
we can combine this with the arclength constraint (\ref{arc}) to get a pair of differential equations for $R$ and $\Theta$ 
describing a rotationally symmetric MOTS
\begin{align}
\ddot{R} = & \frac{2(1-\dot{R}^2)}{R} - \frac{\dot{R} \sqrt{1-\dot{R}^2}}{R} \cot \Theta\label{ddR}  \\
& + \frac{1}{2R} \sqrt{ \frac{1 - \dot{R}^2}{1-F}} \left( \left(R F' + 2 (1-F)\right) \dot{R}^2 - 4 (1-F)\right) \nonumber 
\end{align}
and
\be
 \dot{\Theta} = \frac{2 \pi  \sqrt{1-\dot{R}^2}}{R} \, ,  \label{dT}
\ee
where we have assumed that $\dot{\Theta} > 0$ (which turns out to be true for all the situations in which we are interested). 

We need initial conditions in order to solve these equations. By the assumed symmetry if we choose $\lambda=0$ at $\theta=0$ (the north pole)
then 
\be
\dot{R}(0) = 0 \, . \label{IC1}
\ee
Thus given a choice
\be
R(0) = R_o \label{IC2}
\ee
for some constant $R_o$ we can find a MOTS candidate. These equations can always be integrated and so by construction will always produce a $\tl = 0$ 
surface. Its closure or lack thereof will determine whether or not it is a MOTS. This shooting method is commonly used for finding axisymmetric apparent horizons in 
numerical relativity\cite{Baumgarte:2010:NRS:2019374,Thornburg2007}. 
%
%

\subsection{Numerical Examples}

\begin{figure}
\includegraphics[scale=0.48]{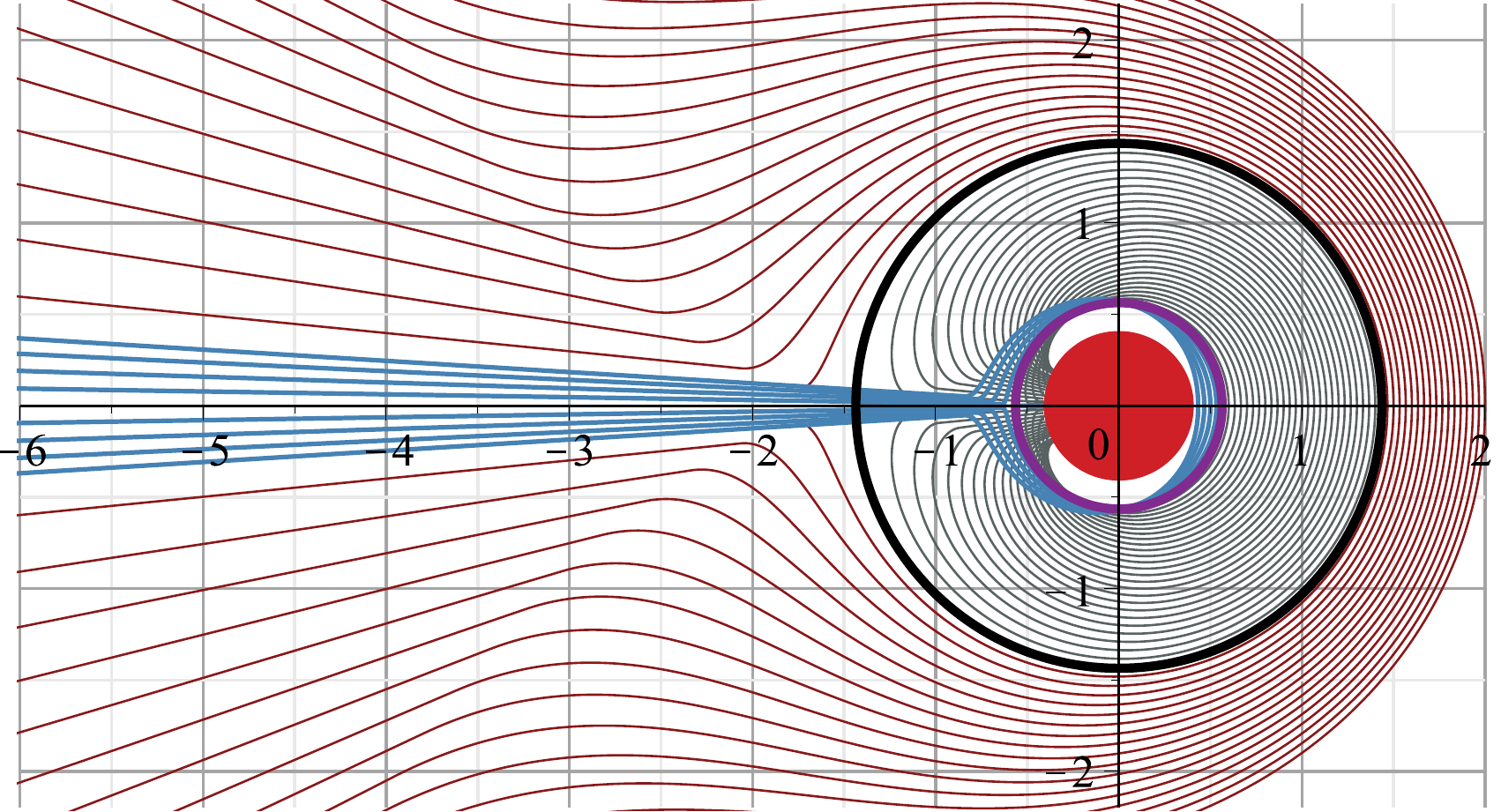}
\caption{Axi-symmetric 
$\tl=0$ surfaces in a $q/m = 0.9$ Reissner-Nordstr\"{o}m spacetime. The inner horizon is purple and the outer horizon is black. Other (open) $\tl=0$ surfaces are numerically
solved from initial conditions (\ref{IC1}) and (\ref{IC2}) and coloured blue, grey or dark red  depending on the value of $R_o$. The red circle in the middle is the region that is
not covered by the PG coordinates. The $z$-axis is horizontal with the north pole on the right-hand side. }
\label{Fin_RN}
\end{figure}

\begin{figure}
\includegraphics[scale=0.48]{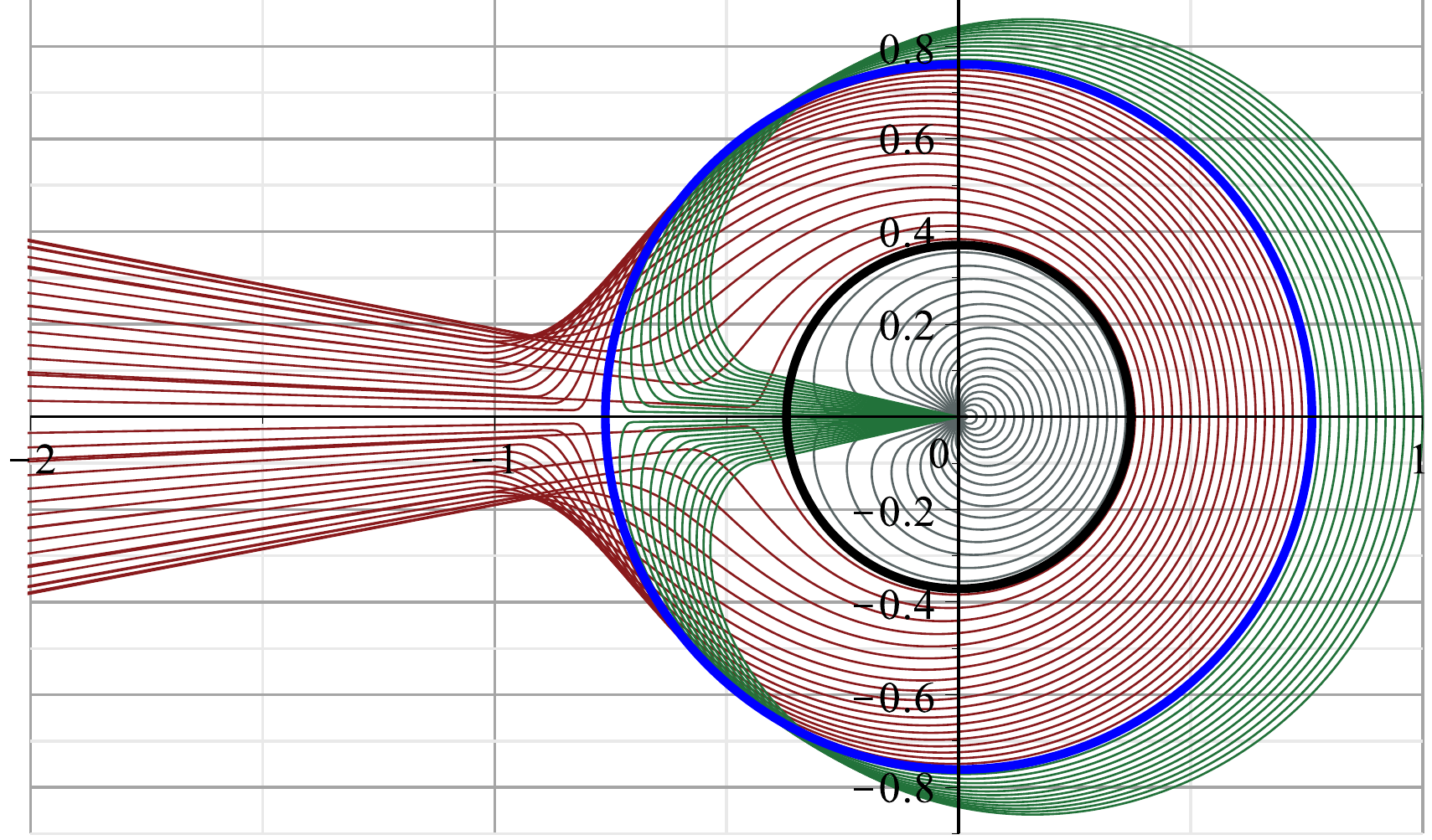}
\caption{Axi-symmetric $\tl=0$ surfaces in $\Lambda=0.0768/m^2$ Schwarzschild-deSitter spacetime. The outer black hole and cosmological horizons are respectively black and blue. 
Other (open) $\tl=0$ surfaces are numerically solved from initial conditions (\ref{IC1}) and (\ref{IC2}) and grey, dark red or green depending on the value of 
$R_o$. The $z$-axis is horizontal with the north pole on the right-hand side. } 
\label{Fin_SdS}
\end{figure}

Some sample $\tl = 0$ surfaces are shown in FIG.~\ref{Fin_RN} and FIG.~\ref{Fin_SdS} which respectively show typical Reissner-Nordstr\"om and Schwazschild-deSitter spacetimes.
Those figures show the system (\ref{ddR})-(\ref{dT}) solved with initial conditions (\ref{IC1}) and (\ref{IC2}). Solutions were obtained using Maple's\cite{maple} built-in routines for 
systems of differential equations. Note that while the known horizons certainly show up as solutions there is also a $\tl = 0$ surface running through all points on the positive $z$-axis. 
The behaviours shown in the figures appear to be generic. Axisymmetric $\tl =0$ surfaces that originate from $0 < R_o < r_{\mbox{\tiny{IH}}}$ and 
$r_{\mbox{\tiny{OH}}}<R_o<r_{\mbox{\tiny{CH}}}$ ultimately diverge to infinity while those from $r_{\mbox{\tiny{IH}}} < R_o < r_{\mbox{\tiny{OH}}}$ and 
$r_{\mbox{\tiny{CH}}}<R_o<\infty$ turn in and disappear into the singularity (or in the case with $q\neq0$ the region where the coordinate system is no longer defined). 
Thus those surfaces are not MOTS as they are not smooth and closed.

These divergences can be contrasted with the now familiar pure deSitter case. For that spacetime 
\be
\theta_{\hat{u}} = - 2 \sqrt{ \frac{3}{\Lambda} } 
\ee
and so in a $T=\mbox{constant}$ surface any sphere of radius 
\be
r_S = \sqrt{\frac{3}{\Lambda}} 
\ee
will have $\tl = 0$. Examples are shown in FIG.~\ref{Fin_RN} (which despite the preceding analysis were numerically evolved in the same way as the previous examples). 
\begin{figure}
\includegraphics[scale=0.45]{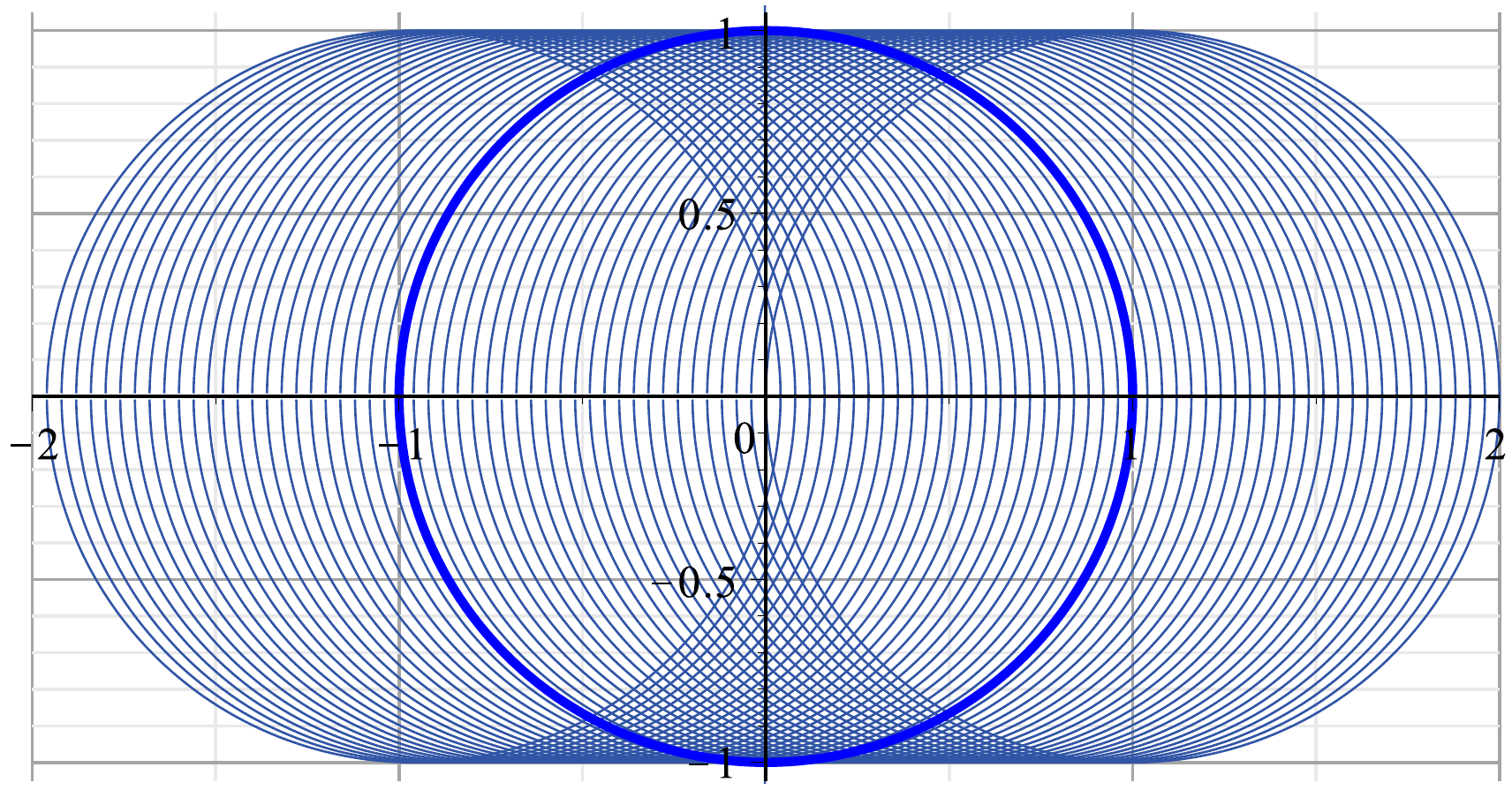}
\caption{Translated cosmological horizons in pure deSitter spacetime. Coordinates are in terms of the deSitter radius. }
\label{Fin_dS}
\end{figure}

Note  that the open $\tl =0$ surfaces  shown in FIG.~\ref{Fin_RN} are not leaves of an isolated horizon\cite{Ashtekar:2004cn}. That is, if a particular $\tl = 0$ surface that opens up to infinity
(or plunges into the singularity) is extended to a three-surface as the locus of points it traces as $T$ varies, then that three-surface is not null. So 
while these are $\tl = 0$ surfaces, they should not be viewed as foliating a kind of ``open'' horizon.

As a side note, the ubiquity of $\tl = 0$ surfaces seen in  FIG.~\ref{Fin_RN} and FIG.~\ref{Fin_SdS} serves to emphasize the non-local character of MOTS: finding a $\tl = 0$ surface is not difficult and in 
our examples it is possible to find such a surface through any point. The hard part is finding a $\tl=0$ that smoothly closes. Determining whether or not that happens requires an integration to find
the full surface. Hence whether or not a particular section of a $\tl =0$ surface is  part of a MOTS may be determined by the detailed geometric properties of a far-away section of spacetime.

\subsection{Higher order stability}

We now return to the first-order unstable cases found in Section \ref{InfDef} to investigate their stability at higher order. 

We begin with numerical tests: evolving from initial conditions
$R_o = \rIH + \delta R_o$ where $\rIH$ is the MOTS of an inner Reissner-Nordstr\"{o}m horizon while $\delta R_o$ is a finite perturbation. 
Finite instabilities will manifest as finite deformations (like those in FIG.~\ref{Fin_dS}) while higher order stability will mean that any initially finite deformation will diverge 
(like those in FIG.~\ref{Fin_RN} and FIG.~\ref{Fin_SdS}).  

\begin{figure*}
\includegraphics[scale=.85]{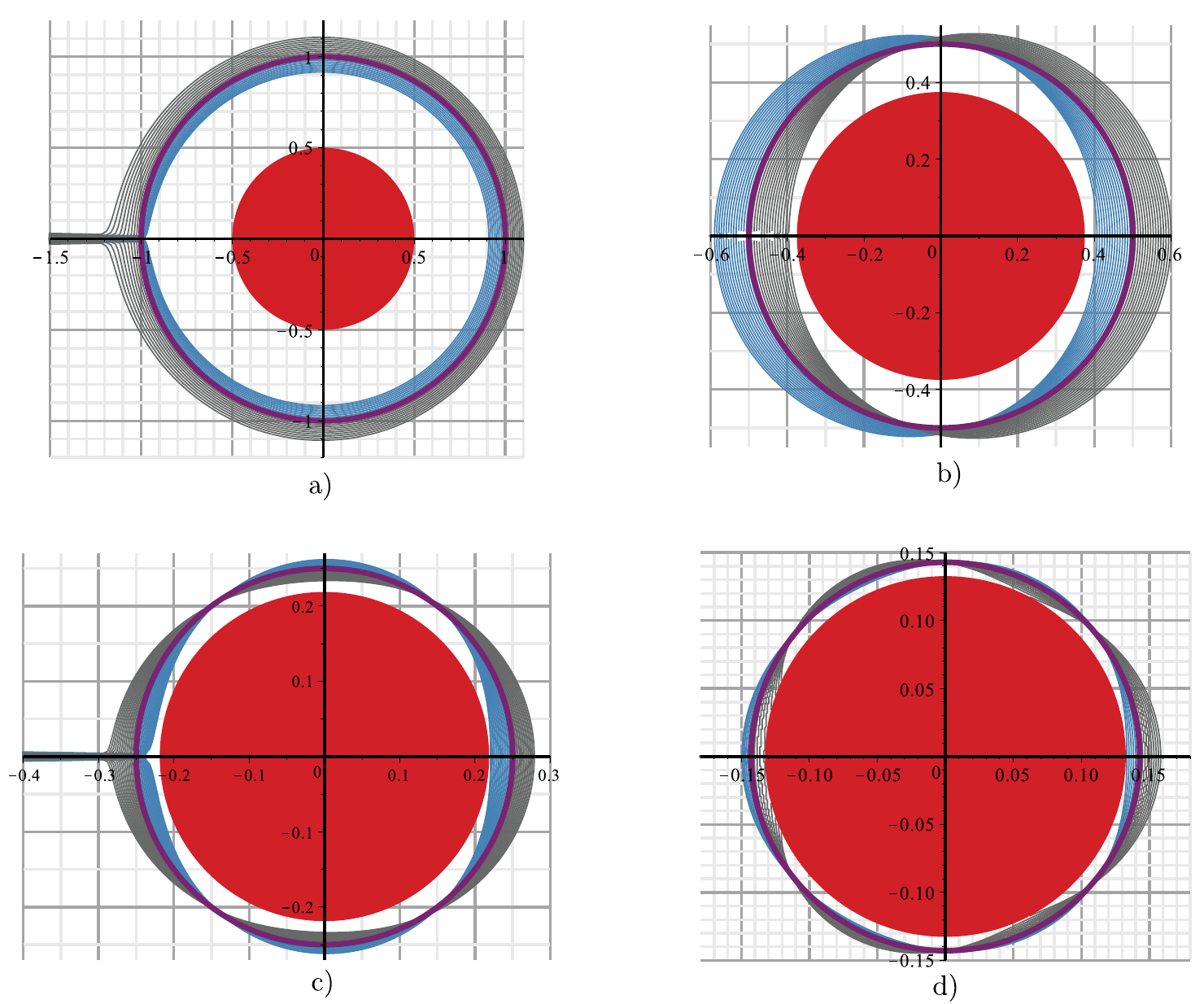}
\caption{Numerical solutions of first-order unstable inner Reissner-Nordstr\"om horizons for a) $l=0$, b) $l=1$, c) $l=2$ and d) $l=3$. $l=0,2$ appear to diverge for all sizes of 
initial perturbation but no divergence can be seen for $l=1,3$.  As in FIG.~\ref{Fin_RN} the red centre is not covered by the PG coordinates. } 
\label{Testing_RN}
\end{figure*}

The $l=0,1,2,3$ modes are shown in FIG.~\ref{Testing_RN}. For $l=0,2$ the instability appears to fail as the numerical solutions diverge at $\theta = \pi$ however for $l=1,3$ 
things are not so clear. In those two cases we don't observe any divergences. However,  while suggestive, these observations aren't conclusive as in both cases the 
behaviour could change for sufficiently small $\delta R_o$. 

To better understand what is happening we turn to a higher-order analysis of the equations. Then it is more 
convenient to work with a single function and so we 
switch to parameterize $R$ with $\theta$. Working from
\be
\lambda_\theta = \sqrt{R_\theta^2 + R^2}
\ee
and 
\be
\dot{R} =\frac{R_\theta}{\lambda_\theta}
\ee
(where derivatives with respect to $\theta$ are indicated with subscripts)
the conversion is straightforward and we get the following differential equation:
\begin{align}
0 =&  R_{\theta \theta}  -  \left(\frac{3}{R} + \frac{\sqrt{R^2 + R_\theta^2}}{2R^2 \sqrt{1-F}} \left(RF'- 2(1-F) \right)\right) R_\theta^2  \nonumber \\
& +  \left(\frac{ R^2 + R_\theta^2  }{R^2}\right)  \cot \! \theta R_\theta  - 2R     + 2 \sqrt{(1-F)( R^2 + R_\theta^2)} \, . \label{Rtt}
\end{align}
Given an inner horizon at $r_{\mbox{\tiny{IH}}}$ we can then look for MOTS of the form:
\be
R(\theta) = r_{\mbox{\tiny{IH}}} + m (  \epsilon R_1 (\theta) + \epsilon^2 R_2 (\theta) + \epsilon^3 R_3 (\theta) + \dots ) 
\ee
where of course $\epsilon$ is the initial perturbation from $\rIH$ at $\theta=0$. Then we have initial conditions
\be
R_1 (0)  = 1 \; \mbox{ and } R_L(0) = 0 \; \mbox{ for } L>1 
\ee
while the first derivative $R_{L\theta} (0) = 0$ for all $L$.

We also expand $F = 1 - \frac{2m}{r} + \frac{q^2}{r^2}$ as a Taylor series around $\rIH = m - \sqrt{m^2 - q^2}$ as:
\be
F(r) = \sum_{n=1}^{\infty} (-1)^n\big(n l (l+1)  +(n-1) \big) \left(  \frac{r- \rIH}{\rIH} \right)^n
\ee
where deriving these expansion uses the first-order condition $\rIH F_1 = - l(l+1)$.

%

Then to the first three orders (\ref{Rtt}) expands as
\begin{align}
 0 = & \triangle_l R_1 \label{R1} \\
 0 = & \triangle_l R_2  + \left(  \frac{l^2(l+1)^2 - 4  }{4}   \right) R_{1\theta}^2 \label{R2} \\
 &   - \left(\frac{(l^2 + l + 2)^3}{8} \right) R_1^2 \nonumber  \\
 0 = & \triangle_l R_3  + \left( \frac{(l^2+l+2)^2 \cot \theta}{4} \right) R_{1\theta}^3 \\ 
  & - \left(\frac{l(l+1)(l^2+l+2)^2(l^2+l+4)}{16}\right) R_1 R_{1 \theta}^2 \nonumber \\
  & + \left(\frac{l^2(l+1)^2 - 4}{2} \right) R_{1 \theta} R_{2 \theta} \nonumber \\
  & - \left(\frac{(l^2+l+2)^3}{4} \right) R_1 R_2 + \left(\frac{(l^2+l+2)^5}{32} \right) R_1^3 \label{R3}
  \nonumber
\end{align}
where $\triangle_l$ is the second-order differential operator that vanishes for first-order unstable perturbations (\ref{SL}):
\be
\triangle_l = \frac{d^2}{d\theta^2} + \cot \! \theta \frac{d}{d\theta} + l (l+1)  \, .  \label{trianglel}
\ee
In the following we refer to the non-$\triangle_l R_n$ terms in each equation as $h_{ln}$. 

Of course the $R_1$ equation is (\ref{SL}) again. Note that the equations can be solved sequentially. Once we have $R_1$ we can solve for $R_2$ and then both of them 
can be used to solve for $R_3$. While the rapidly growing complexity  of the expressions means that it is not practical to show the higher order equations, this pattern continues. 

%
%
%

Then we are interested in solutions to equations of the form 
\be
\triangle_l X + h(\theta) = 0 \, . \label{Prob}
\ee 
where $X(\theta):[0,\pi] \rightarrow \mathbb{R}$ satisfies initial conditions $X(0)=X_o \in \mathbb{R}$, $X_\theta (0) = 0$ and
$h(\theta)$ can be expressed as a finite sum of Legendre polynomials:
\be
h (\theta) = \sum_{L=0}^{L_{\mbox{\tiny{max}}}} h_{[L]} P_L (\cos  \theta) \, .
\ee

Now, the solution to the homogeneous version of (\ref{Prob}) is the Legendre polynomial $P_l(\cos\theta)$.  Therefore, by the Fredholm Alternative theorem, the inhomogeneous problem 
has a solution if and only if $h(\theta)$ is orthogonal to $P_l(\cos\theta)$: 
\be
h_{[l]} \equiv \frac{2l+1}{2} \int_0^\pi \! \!  \sin  \theta \, h(\theta) P_l (\cos  \theta)  d  \theta = 0 \, . \label{htest}
\ee 
Equivalently the Legendre polynomial expansion of $h$ does not contain an $h_l$ term. 

If a solution does exist then it is also a finite sum of Legendre polynomials:
\be
X (\theta) = \sum_{L=0}^{L_{\mbox{\tiny{max}}}}X_L P_L (\cos  \theta) \label{X}
\ee
where for $L \neq l$
\be
X_L = \frac{h_{[L]}}{L(L+1)-l(l+1) } \label{X1}
\ee
and 
\be
X_l = X_0 - \left. \sum_{L=0}^{L_{\mbox{\tiny{max}}}} \right|_{L \neq l} X_L \, . \label{X2}
\ee

%
%
%
%
%

These observation can then be combined as an algorithm to test for solutions to the deformation problem to arbitrary order. 
\begin{enumerate}
\item Set $R_1 = \epsilon m  P_l (\cos \theta)$ and $n=2$. 
\item \label{s2} Find the $\epsilon^n$ term in the expansion of (\ref{Rtt}).
It will take the form
\be
\triangle_l R_n + h_{ln}  = 0 \label{heq}
\ee
where  $h_{ln}$ will always be a sum of terms involving $l$, $\cot \theta$, $R_1$, $R_{1\theta}$, $R_2$, $R_{2\theta}$, \dots, 
$R_{n-1}$ and $R_{(n-1) \theta}$.
\item Substitute the known expressions for $R_m$ into $h_{ln} (\theta)$, $m<n$ and 
use (\ref{htest}) to test whether (\ref{heq}) has a divergent solution. If $h_{ln[l]} \neq 0$, stop. This case 
 cannot be finitely deformed. 
\item If the solution of (\ref{heq}) isn't divergent, use (\ref{X})--(\ref{X2}) to generate $R_{n+1}$ and repeat from Step \ref{s2} with $n \rightarrow n+1$.  
\end{enumerate}
Appendix \ref{triple} recalls  some results on series expansions of derivatives and products of Legendre polynomials that are used in implementing these steps. 

For a finite deformation this algorithm would never terminate. However in all cases that we have checked
we find a divergence at some order and so the horizon cannot be finitely deformed. 

\begin{figure}
\includegraphics[scale=0.5]{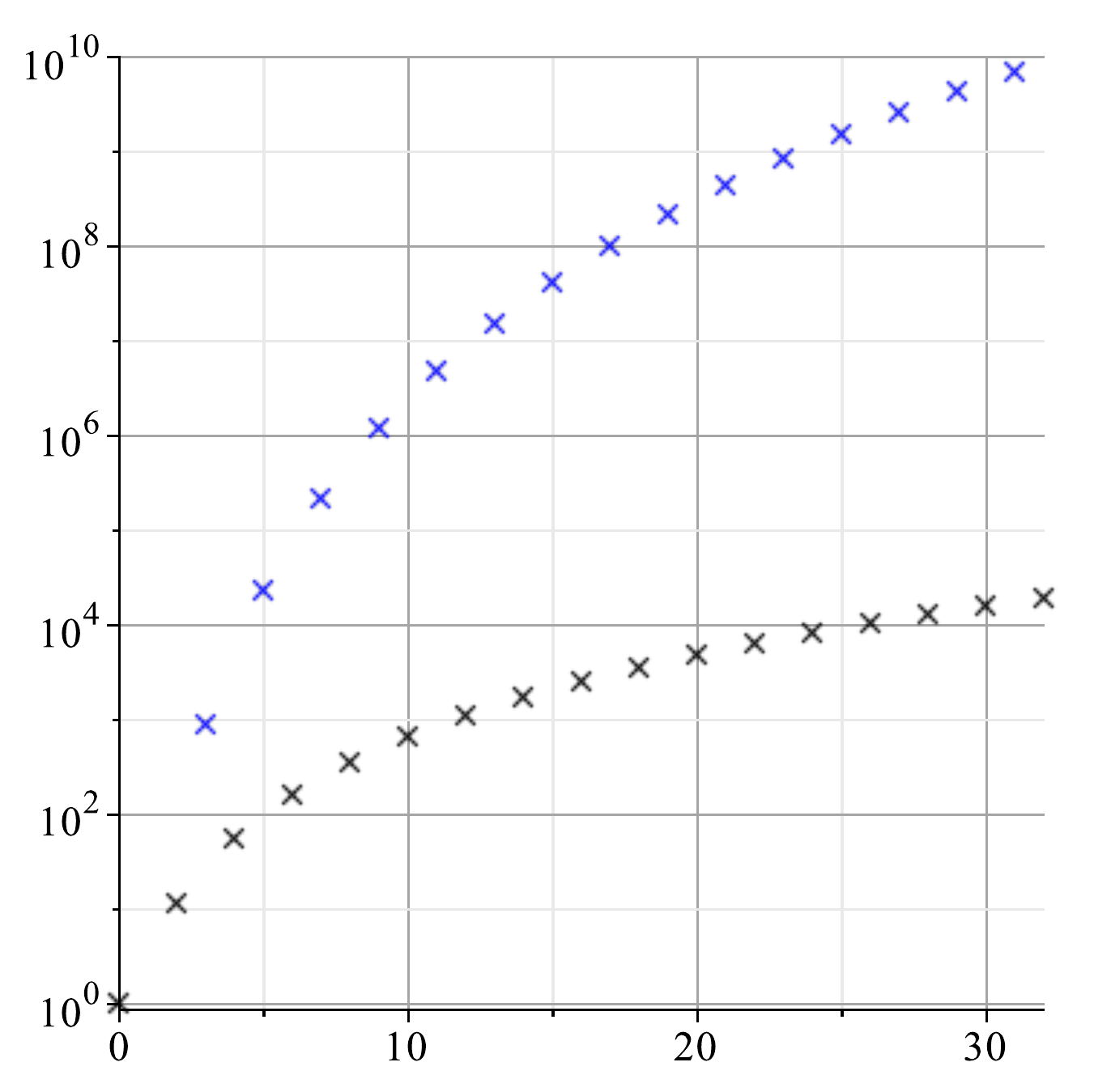}
\caption{$-h_{l2[l]}$ (black) and $h_{l3[l]}$ (blue) for even and odd cases respectively for $l=2$ to $l=31$. Both appear to be growing monotonically. }
\label{EvenOddPlot}
\end{figure}

Explicitly, the first four even $l$ cases are
\begin{align}
\left. \frac{R}{m} \right|_{l=0} \mspace{-16mu} & \approx 1+ P_0  \epsilon+ \mbox{(divergent term)} \epsilon^2\\
\left. \frac{R}{m} \right|_{l=2} \mspace{-16mu} & \approx  \frac{1}{4} + P_2  \epsilon+ \mbox{(divergent term)} \epsilon^2 \\
\left. \frac{R}{m} \right|_{l=4} \mspace{-16mu} & \approx \frac{1}{11} + P_4   \epsilon+ \mbox{(divergent term)} \epsilon^2 \\
\left. \frac{R}{m} \right|_{l=6} \mspace{-16mu} & \approx \frac{1}{22} + P_6   \epsilon + \mbox{(divergent term)} \epsilon^2  \, , 
\end{align}
while the first three odd cases are 
\begin{align}
\left. \frac{R}{m} \right|_{l=1} \mspace{-16mu}  \approx & \frac{1}{2} + P_1 \epsilon+ 4 \left( \frac{P_0-P_2}{3} \right) \epsilon^2  +  16 \left( \frac{-P_1 + P_3 }{5} \right)  \epsilon^3 \nonumber \\
& +  8  \left( - \frac{17}{15} P_0 + \frac{7}{3} P_1 + \frac{1}{21} P_2 - \frac{131}{105} P_4 \right)  \epsilon^4  \\
& + \mbox{(divergent term)} \epsilon^5 \nonumber  \\
\left. \frac{R}{m} \right|_{l=3} \mspace{-16mu} \approx & \frac{1}{7} + P_3  \epsilon + \bigg(  -\frac{11}{12} P_0 + \frac{8}{9} P_2 + \frac{35}{2} P_3 \\
& \mspace{30mu}  - \frac{351}{44}  P_4 - \frac{940}{99} P_6 \bigg) \epsilon^2  + \mbox{(divergent term)} \epsilon^3 \nonumber\\
\left. \frac{R}{m} \right|_{l=5}  \mspace{-16mu} \approx & \frac{1}{16}  + P_5 \epsilon+ \bigg( - \frac{1312}{165} P_0 - \frac{12200}{1287} P_2 - \frac{3456}{715} P_4 \nonumber \\
& + 112 P_5 - \frac{41600}{1683} P_6 - \frac{190400}{8151} P_8 - \frac{9620856}{230945} P_{10} \bigg) \epsilon^2 \nonumber \\
&  + \mbox{(divergent term)} \epsilon^3  \; . 
\end{align}
In these expressions the  $\cos \theta$ dependence of the $P_n$ is suppressed.  The pattern  appears to continue for all $l>1$: that is even $l$ diverge at second order while 
odd cases diverge at third order. This is demonstrated in FIG.~\ref{EvenOddPlot} where  $h_{l2[l]}$ and $h_{l3[l]}$ are plotted up to $l=31$. 

Though the trend in FIG.~\ref{EvenOddPlot} seems clear we were not able to show that the growth continues for all $l$: the expressions, particularly for the odd cases, become 
prohibitively complex. 

However we can at least see why the odd cases don't diverge at second order. Reading off $h_{2l}$ from (\ref{R2})
and keeping in mind that $R_{1l} = P_l (\cos \theta)$ then applying (\ref{prodEx})  it is straightforward to see that on expanding the  $R_1^2$ term into a Legendre series
we only obtain even terms. Meanwhile the $R_{1\theta}^2$ term can be expanded using (\ref{Pp}) and then (\ref{prodEx}) again to see that it also contains
only even terms. Hence $h_{2l[l]}$ necessarily vanishes for all odd $l$ and any divergence must be at third or higher order.

\section{Discussion}
\label{Discuss}

While a stable MOTS cannot be smoothly deformed, ``instability'' is not sufficient to imply that such deformations are possible. For spherically symmetric MOTS in similarly symmetric spacetimes
we first showed that all but a few finely tuned cases are not deformable. Then on checking those special cases for RN spacetimes we saw that, apart from pure deSitter spacetime, none of them
appeared to be deformable either (though we did not find a completely general proof). 

Hence we expect that a much stronger result holds that prevents (virtually all) horizons from being deformed. We expect that only spacetimes with extra symmetries (such as 
pure deSitter) can house deformable MOTS and that those deformations will turn out to be translations. 

This would not be a particularly shocking result. It would be much more surprising to discover that inner black hole MOTS could be finitely deformed. However as 
far as we know there is no extant theorem that proves this.

\acknowledgements

This work was supported by NSERC Discovery Grants 261429-2013 (IB)  and 418537-2012 (HK). AO was also supported by 
an NSERC Undergraduate Student Research Award and a Memorial University Student Internship Award.

%

%
%
%

%
%



\appendix

\section{Useful Legendre Identities}
\label{triple}
In this appendix we recall how to expand the derivative and product of Legendre polynomials as a Legendre series. These are used in solving the horizon deformation equations. 

\subsection{Derivatives}
By the standard recurrence relations it is straightforward to show that the derivative of a Legendre polynomial can be expanded as
\be
 P_l' (\cos \theta)  = \left\{ 
\begin{aligned}
&\sum_{m=1}^{\nicefrac{l}{2}} (4m-1) P_{2m-1}(\cos \theta) & l \mbox{ even}\\ 
&\sum_{m=0}^{\nicefrac{(l-1)}{2}} (4m+1) P_{2m}(\cos \theta) & l \mbox{ odd} 
\end{aligned} 
\right\}  \, .  \label{Pp}
\ee
where $P'_l (x) = {dP_l(x)}/{dx}$.

\subsection{Products}
Next recall that the integral of three polynomials $P_k(x)$, $P_l(x)$ and $P_m(x)$ is given by the Wigner 3j symbol: 
\begin{align}
\int_{-1}^{1} P_k P_l P_m dx = 2 \left( \begin{array}{ccc} k & l & m \\ 0 & 0 & 0 \end{array} \right)^2 
\end{align}
where if $|k-l|\leq m \leq k+l$ and $2s = m +k +l$ is even then
\begin{align}
\left( \begin{array}{ccc} k & l & m \\ 0 & 0 & 0 \end{array} \right)^2 = & \;  \frac{(2s-2k)!(2s-2l)!(2s-2m)!}{(2s+1)!} \\
& \times  \left( \frac{s!}{(s-k)!(s-l)!(s-m)!}\right)^2 \nonumber
\end{align}
else it is zero. 

Then we can series expand the product of polynomials as a finite series
\be
P_k P_l = \sum_{m=|k-l|}^{k+l} (2m+1) \left( \begin{array}{ccc} k & l & m \\ 0 & 0 & 0 \end{array} \right)^2 P_m \, . \label{prodEx}
\ee
For purposes of the discussion in the main text the important point is that for $k+l$ even there are only even terms in the expansion while for 
$k+l$ odd there are only odd terms.

\bibliography{DistortingMOTS}

\end{document}